\documentclass[preprint,12pt]{elsarticle}
\usepackage{mathrsfs}
\usepackage{amssymb}
\usepackage{amsbsy}

\usepackage{amssymb}

\journal{Physics Letters A}

\begin{document}

\begin{frontmatter}



\title{Two-component feedback loops and deformed mechanics}


\author{David S. Tourigny}

\address{D.A.M.T.P., Centre for Mathematical Sciences, University of Cambridge, Wilberforce Road, Cambridge CB3 0WA, U.K.}

\begin{abstract}
It is shown that a general two-component feedback loop can be viewed as a deformed Hamiltonian system. Some of the implications of using ideas from theoretical physics to study biological processes are discussed.
\end{abstract}


\end{frontmatter}


\section{Introduction}

The physicist Eugene Wigner famously wrote on the enormous contribution that mathematical theories have made to physics \cite{Wig}. By contrast, mathematics has so far had less impact on biology \cite{Les}. Biological systems can be viewed as an emergence of the laws of chemistry and the principle of natural selection, and this underlying complexity makes biological processes incredibly difficult to study mathematically. There are many instances where symmetry apparent on the level of an organism breaks down when one is to look on the molecular or cellular scale. For example, the body plans of most animals display some form of radial or bilateral symmetry, but this is not a symmetry in the exact sense as is revealed if one is to consider the arrangement of cells. Consequently, the symmetry of an organism can only be considered a symmetry `from far away' much like the symmetries of statistical phenomena or the apparent homogeneity of the cosmos.

On the other hand, more exotic symmetries can be found in the elementary world described by the standard model of particle physics. There the concept of spontaneous symmetry breaking is well understood to play a role in hadron formation, but complications immediately arise when one is moved to consider interactions between multiple hadrons and the higher atomic nuclei. Thus, the problem of describing biological processes mathematically seems to be associated with the problem of symmetry. Whilst mathematics describes well the physics of the very large and very small it does not appear to cope well with molecular or cellular biology, trapped, unsymmetrically, with the insufficiently large and the insufficiently small.    

In classical mechanics, Noether's theorem states that for every symmetry of the equations of motion there is a corresponding conserved quantity. By a conserved quantity is meant a function of the dynamical variables that does not vary in time so that its total time derivative always remains zero. In Hamilton's formulation of classical mechanics it is the Hamiltonian representing the total energy of the system that always remains conserved, but in a typical biological process there is no analogue of the Hamiltonian and therefore no conserved quantity. Exceptions to this rule emerge when the equations governing these dynamical systems can be put into Hamiltonian form. This has been achieved for the classical Lotka-Volterra equations that govern predator-prey interactions \cite{Nut,Pla}, and simple signalling models involving constant degradation rates \cite{Goo}. In these cases an analogue of the Hamiltonian  immediately yields a conserved function of the dynamical variables that can be used to study Lyapunov stability and the location of equilibrium points. 

In physics, scale invariance is a feature of equations or observables that does not change if the scales of certain variables are multiplied by a common factor (often forming part of a larger conformal symmetry). Scale-invariance is a typical property of critical phenomena because experimental observables are known to follow power-laws near the neighbourhood of a critical point. There is now a growing realisation that scale-invariance may be an inherent feature of many biological networks that display critical behaviour (this notion of scale-invariance is distinct from that of a network being scale-free, a topological property) \cite{Gis}. For example, recent work demonstrates that pathways involved in growth factor signalling are dependent on fold-changes in concentrations of a molecule, and not its absolute level \cite{Goe}. This is an indication that certain biological processes display at least some evidence of a symmetry. 

In this paper it is suggested that many scale-invariant biological processes can be viewed as a deformation of classical mechanics. In particular, the focus is on cellular signalling pathways and changes in the concentrations of their constitutive molecules. Here the symmetries present in the Hamiltonian formalism are deformed in a known way (that reflects deviation of the underlying system from being conservative) and so a conserved quantity can be recovered after reversing the deformation. This reversal relies heavily on the use of $q$-calculus, which is a common feature of deformed mechanics. The models considered have very general applications and the procedure for passing to conservative dynamics from a deformed system is likely to emerge as a common tool for studying near-symmetric biological processes. A detailed account of $q$-deformed mechanics is provided in Section 2 after a review of classical dynamics for readers not familiar with the Hamiltonian formalism. The general model and an illustrative example is considered in Section 3 prior to a higher-dimensional extension in Section 4. 

\section{Review of $q$-deformed classical dynamics in one dimension}      
In the Hamiltonian formulation of one-dimensional classical mechanics it is convenient to consider a two-dimensional real vector space $V$ isomorphic to $\mathbb{R}^2$. This vector space is usually called phase space and the two components $(x^1,x^2)$ of a vector field $x \in V$ are often referred to as position and momentum coordinates. The phase space $V$  becomes a symplectic vector space when equipped with an anti-symmetric, non-degenerate bilinear form $\Omega: V \times V \to \mathbb{R}$. Specifically, for any $x,y \in V$ the symplectic form $\Omega$ satisfies $\Omega(x,y)=-\Omega(y,x)$ and the feature that if $\Omega(x,y) =0$ for all $y \in V$ then $x=0$. Viewed as a matrix acting on $\mathbb{R}^2$, $\Omega$ can be chosen to have the representation

\begin{equation} 
\Omega = \left( \begin{array}{cc} 0 & 1 \\ -1 & 0  \end{array} \right) \ . 
\end{equation}
The symplectic form defines the Poisson bracket of two functions of the position and momentum coordinates. Writing the components of $x\in V$ as $x^a$ ($a=1,2$) and the components of the symplectic form as $\Omega_{ab}$ (its inverse as $\Omega^{ab}$), the Poisson bracket $\{f,g\}$ of any two functions $f,g: (x^1,x^2) \to \mathbb{R}$ is defined to be the function

\begin{equation}
\{f,g\} = \sum_{a,b} \Omega^{ab} \partial_a f \partial_b g \ ,
\end{equation}
where the operator $\partial_a$ denotes partial differentiation with respect to the coordinate $x^a$. 

In a classical physical system the total energy is a function $H: (x^1,x^2,t) \to \mathbb{R} $ called the Hamiltonian. Consider the total time derivate of a Hamiltonian that does not depend explicitly on time $t$:

\begin{equation}
\dot{H} = \sum_a \partial_a H \ \dot{x}^a = \partial_1H \ \dot{x}^1 + \partial_2 H \ \dot{x}^2 \ .
\end{equation} 
Since the system is not exchanging energy with its environment, the total energy should remain constant over time. Imposing the condition $\dot{H} = 0$ yields the relations $ \partial_1H = \dot{x}^2$ and $ \partial_2H = -\dot{x}^1$, which are precisely Hamilton's equations of motion. These can be written succinctly in terms of the inverted symplectic form 

\begin{equation}
\dot{x}^a = \sum_b \Omega^{ab} \partial_b{H} \ .
\end{equation}   
Once the Hamiltonian has been specified the system is determined uniquely since calculating the total time derivative of any function $f: (x^1,x^2) \to \mathbb{R}$ one can show

\begin{equation}
\dot{f} = \{H,f\}  \ .
\end{equation}   
For example, Hamilton's equations of motion are recovered taking $f$ to be either of the $x^a$.

There is a nice symmetry of Hamilton's equations that can be most easily verified using the Poisson bracket formalism. Consider a coordinate transformation $x^a \to X^i(x^1,x^2)$  ($i=1,2$) with associated Jacobian matrix $J$. The resulting change in the Poisson bracket is

\begin{equation}
\{f,g\} =\sum_{a,b} \sum_{i,j}  \Omega^{ab} J^i_a J^j_b \partial_i f \partial_j g  \ ,
\end{equation} 
which means the dynamics are only preserved if $J$ satisfies

\begin{equation}
J \Omega J^T = \Omega  \label{symplectic}
\end{equation}
so that

\begin{equation}
\dot{X}^i = \sum_j \Omega^{ij} \partial_j{H} \ .
\end{equation} 
Elements of the general linear group $GL(2,\mathbb{R})$ of non-singular $2 \times 2$ real matrices that satisfy (\ref{symplectic}) form a subgroup called the symplectic group $Sp(2,\mathbb{R})$. In two dimensions the symplectic group is isomorphic to the group of non-singular $2 \times 2$ real matrices with unit determinant $SL(2,\mathbb{R})$. Consequently, Hamilton's equations remain invariant under a change of coordinates whose Jacobian is a member of $Sp(2,\mathbb{R})$. A transformation of this kind may also be called canonical since time evolution can be viewed as a one-parameter family of these generated by the Hamiltonian and taking $x^a(0)$ to $x^a(t)$.    

The importance of symmetry is best demonstrated by Noether's theorem that states for every symmetry of Hamilton's equations there is an additional function of the $x^a$ that is conserved in time. To see this, for an infinitesimal transformation 

\begin{equation}
x^a \to x^a + \epsilon F^a(x^1,x^2) + O(\epsilon^2)
\end{equation}   
to be a symmetry the requirement (\ref{symplectic}) demands

\begin{equation}
\partial_1 F^1 = - \partial_2 F^2 \ ,
\end{equation}
which is satisfied if $F^1 = \partial_2 G$ and $F^2 = \partial_1 G$ for some function $G: (x^1,x^2) \to \mathbb{R}$. Then the Hamiltonian transforms infinitesimally as 

\begin{equation}
H \to H + \epsilon \{H,G\}  + O(\epsilon^2) \ ,
\end{equation}
but since the coordinate change is a symmetry of the Hamiltonian

\begin{equation}
0 = \epsilon \{H,G \} = \epsilon \dot{G}
\end{equation}
and therefore $G$ is also conserved. The existence of conserved quantities is useful for qualitative study of complicated isolated systems, but these are rarely present in biological processes operating far from equilibrium.    
 
Deformations of Hamilton's equations arise when one moves to a non-commutative setting, the standard example being canonical quantisation in quantum mechanics. This deformation is on the scale of Planck's constant $h$ and involves promoting $x^1$ and $x^2$ to operators $\bar{x}^1$ and $\bar{x}^2$ that satisfy 

\begin{equation}
\bar{x}^1 \bar{x}^2 - \bar{x}^2\bar{x}^1 = i \hbar \ ,
\end{equation}
where $i=\sqrt{-1}$ and $\hbar= h/2\pi$. A less-well-known example is the $q$-deformation, which involves some fixed real parameter $q$ different than $1$ and takes the form

\begin{equation} \hat{x}^1 \hat{x}^2- q\ \hat{x}^2\hat{x}^1 = 0\ .
\end{equation} 
The variables $\hat{x}^1$ and $\hat{x}^2$ generate a different algebra to that of the real numbers that is referred to as the quantum plane $\mathbb{R}_q^2$. Partial derivative operators of the quantum plane have been defined to fulfil the relation $\hat{\partial}_a \hat{x}^b = \delta^b_a$ ( $\delta^b_a$ is the Kronecker delta) so that they satisfy the $q$-commutative rule

\begin{equation}
\hat{\partial_2}\hat{\partial_1} = q^{-1} \hat{\partial_1}\hat{\partial_2} \ ,
\end{equation}
and various $q$-Leibniz rules that reduce to the calculus of commutative variables in the limit $q \to 1$  \cite{Wes}. 

The most general function on $\mathbb{R}^2_q$ can be expanded in terms of the $\hat{x}^a$ as 

\begin{equation}
\hat{f}(\hat{x}^1, \hat{x}^2) = \sum_{n,m}C_{n m}(\hat{x}^1)^{n} (\hat{x}^2)^{m} \ , \label{expand}
\end{equation} 
for some real numbers $C_{nm}$ that are assumed to absorb the $\hat{x}^1 \leftrightarrow \hat{x}^2$ ordering ambiguity of the polynomial. Then the actions of the $q$-derivatives on the monomials are 

\begin{equation}
\hat{\partial}_1 (\hat{x}^1)^{n} (\hat{x}^2)^{m}= [n] (\hat{x}^1)^{n-1} (\hat{x}^2)^{m}
\end{equation}
and

\begin{equation}
\hat{\partial}_2 (\hat{x}^1)^{n} (\hat{x}^2)^{m}  =[m] q^n (\hat{x}^1)^{n} (\hat{x}^2)^{m-1} \ , 
\end{equation}
where $[n]$ is the $q$-basic number

\begin{equation}
[n] = \frac{q^{2n}-1}{q^2-1}  \ .
\end{equation}
Using these relations it can be confirmed that the rules of ordinary calculus can be regained by working in the limit $q\to 1$. 

In anticipation of what is to come, imagine one is faced with the problem of obtaining a realisation of $\mathbb{R}^2_q$ and its $q$-calculus from the standard commutative algebra of $\mathbb{R}^2$. The solution is to make the identifications
\begin{eqnarray} \label{identi}
\hat{x}^1 \to x^1 \\ \nonumber
\hat{x}^2 \to x^2 \Lambda \\ \nonumber
\hat{\partial}_1 \to \mathcal{D}_1 \\ \nonumber
\hat{\partial}_2 \to \mathcal{D}_2 \Lambda \ ,  \\ \nonumber 
\end{eqnarray} 
where $\Lambda: \Lambda f(x^1,x^2) = f(q x^1,x^2)$ is a dilation or scaling operator acting along the $x^1$-axis and the $\mathcal{D}_a$ defined by

\begin{equation}
\mathcal{D}_1 f(x^1,x^2) = \frac{f(q^2 x^1,x^2)-f(x^1,x^2)}{(q^2-1)x^1} \ , \label{Jack1}
\end{equation} 
and

\begin{equation}
\mathcal{D}_2 f(x^1,x^2) = \frac{f(x^1,q^2 x^2)-f(x^1,x^2)}{(q^2-1)x^2} \ , \label{Jack2}
\end{equation} 
are ``Jackson derivatives'' acting on arbitrary functions $f: \mathbb{R}^2 \to \mathbb{R}$ \cite{Ubr}. Using these identifications, several different $q$-analogues of the Poisson bracket and Hamilton's equations  have been introduced \cite{Lav05,Lav06}.

\section{Two-component feedback loops}
The model considered in this section is a simple feedback loop involving two molecular species $x^1$ and $x^2$:

\begin{eqnarray} \label{network}
\dot{x}^1 =f_1(x^2) - \alpha x^1  \ , \\ \nonumber
\dot{x}^2 = f_2(x^1) - \beta x^2 \ .
\end{eqnarray}
The system (\ref{network}) with constants $\alpha,\beta$ is the most general model involving linear degradation rates and arbitrary activator/repressor terms that do not depend explicitly on the target molecule. By reviewing the relevant literature it can be confirmed that most (if not all) authors have adopted a model of this form when attempting to fit changes in molecular concentrations to experimental data (see, for example \cite{Goo2,deJ,Kar}). In particular, if $x^1$ is chosen to represent the concentration of an mRNA transcript of protein $x^2$, then it is common to assume linear protein expression rates ($f_2(x^1)= \gamma x^1$ with constant $\gamma$) and take $f_2(x^1)$ to be monotonically increasing or decreasing depending on whether $x^2$ is an activator or repressor, respectively. 

Additional molecular species are sometimes incorporated into the system to serve as an intermediary between $x^1$ and $x^2$. Without loss of generality these intermediaries will not be considered here. This is because the concentration of a third molecular component $x^3$ whose dynamical equation does not depend on $x^1$, say, can always be expressed in terms of $x^2$ and substituted into $f_1(x^3)$ to reduce such an intermediary system of three-or-more equations to the form (\ref{network}). A typical example described in this manner is a two-component feedback loop such as the {\em lac} operon that also depends on the concentration of an external compound (e.g. lactose). 

On closer inspection it becomes clear that the system (\ref{network}) can be written
\begin{equation}
\dot{x}^a = \sum_b \Omega_q^{ab} \partial_b H  \ , \label{compact}
\end{equation}
where

\begin{equation}
H = \int f_2 \ dx^1  +\frac{\beta}{\alpha} \int f_1 \ dx^2  - \beta x^1 x^2 \ , \label{pseudoham}
\end{equation}
and $\Omega_q$ is a $q$-deformed symplectic form with matrix representation

\begin{equation} 
\Omega_q =   \left( \begin{array}{cc} 0 &  1 \\   -q & 0 \end{array} \right) \ , \ \ \ q= - \beta/\alpha \ .
\end{equation}
Using a similar argument to that presented in Section 2, it follows that (\ref{compact}) remains invariant under any change of coordinates whose associated Jacobian $J$ satisfies
\begin{equation}
J \Omega_q J^T = \Omega_q  \ .
\end{equation} 
The matrix subgroup $Sp_q(2,\mathbb{R}) = \{A \in GL(2, \mathbb{R}) : A \Omega_q A^T = \Omega_q \}$ is called the $q$-symplectic group. The transformation corresponding to $J$ will be called a $q$-canonical transformation. One consequence is that matrices of the form

\begin{equation}
 \left( \begin{array}{cc} \lambda &  0 \\  0  & \lambda^{-1} \end{array} \right) \ , \ \ \ \lambda \in \mathbb{R} \ ,
\end{equation}
make up a subgroup of $Sp_q(2,\mathbb{R})$ and so systems such as  (\ref{network}) are manifestly scale-invariant (i.e. depends only on the product $x^1x^2$). Indeed, the model used to study the scale-invariant signalling pathway described in \cite{Goe} is just one particular example covered by the generalised system (\ref{network}).

Although $H$ in (\ref{pseudoham}) closely resembles a Hamiltonian, it has no obvious physical interpretation because, unlike its mechanical counterpart, it is not conserved and instead allowed to vary in time. To see this first consider the total time derivative of an arbitrary function $f:(x^1,x^2) \to \mathbb{R}$ in terms of $q$-Poisson bracket

\begin{equation}
\dot{f} = \partial_1f \ \dot{x}^1 + \partial_2 f \ \dot{x}^2 = \sum_{a,b} \Omega_q^{ab} \partial_aH \partial_b f \equiv \{H,f\}_q  \ .
\end{equation} 
Then a simple calculation yields

\begin{equation}
\dot{H} = \{H,H\}_q = (1-q^{-1}) \partial_1H \partial_2 H  \ ,
\end{equation}
which only vanishes in the limit $q \to 1$ (corresponding to $\beta \to -\alpha$). Scale invariance and general $q$-symplectic symmetry suggests existence of some $q$-analogue of Noether's theorem however, and so it may be informative to try and tease out the details.

In Section 2 it was demonstrated how to construct a $q$-deformed version of Hamilton's equations starting from a conservative dynamical system. Is it possible to reverse this procedure and obtain a conservative version of (\ref{compact})? It turns out that this is possible if one is to replace partial derivatives with their Jackson analogues using identifications (\ref{identi}) with $1/q$ in place of $q$. Specifically, if one uses the operators $(\mathcal{D}_1, \mathcal{D}_2\Lambda)$ instead of $(\partial_1, \partial_2)$ it can be shown that there exists some function $H_q$ that remains constant with time. First remark that (\ref{network}) owes the fact it can not be written in standard form to a $x^1 x^2$ cross-term that must appear in the Hamiltonian. Unless $\beta= - \alpha$ then no Hamiltonian can be found to reproduce (\ref{network}) and the distance of $\beta$ from $-\alpha$ must be accounted for by the factor of $q$ appearing in the deformed symplectic form $\Omega_q$. Now instead consider the action of $\mathcal{D}_1$ and $-\mathcal{D}_2\Lambda$ (remembering to use $q=-\alpha/\beta$) on the generalised Hamiltonian

\begin{equation}
H_q = F_2(x^1)-F_1(x^2) -\beta x^1 x^2  \ . \label{qHamil}
\end{equation}  
In particular,

\begin{equation}
\mathcal{D}_1 H_q = \mathcal{D}_1 F_2(x^1) - \beta x^2 \ ,
\end{equation} 
and

\begin{equation}
- \mathcal{D}_2 \Lambda H_q = \mathcal{D}_2 F_1(x^2) +q \beta x^1 = \mathcal{D}_2 F_1(x^2) -\alpha x^1 \ .
\end{equation}
Thus $\mathcal{D}_1H_q=\dot{x}^2$ and $\mathcal{D}_2 \Lambda H_q=-\dot{x}^1$ if and only if the $F_a$ satisfy

\begin{equation}
\mathcal{D}_a{F(x^a)} = f(x^a) \ ,
\end{equation}
which in turn implies (using (\ref{Jack1}) and (\ref{Jack2})) that

\begin{equation}
F_a(x^b) = f_a(x^b)\ x^b \ , \ \ \ (a \neq b)  \ .
\end{equation}
Hence, taking the total time derivative of $H_q$ using $(\mathcal{D}_1, \mathcal{D}_2\Lambda)$ in place of $(\partial_1, \partial_2)$ yields

\begin{equation}
\dot{H}_q = (\mathcal{D}_1 H_q )\dot{x}^1 +( \mathcal{D}_2 \Lambda H_q) \dot{x}^2 = 0 \ .
\end{equation} 

What is the physical interpretation of using $(\mathcal{D}_1, \mathcal{D}_2\Lambda)$ in place of $(\partial_1, \partial_2)$? From the explicit actions of $\Lambda$ and the $\mathcal{D}_a$ it is clear that these operators induce a re-scaling of arguments of the functions they act upon. One can consider the substitution $(\partial_1, \partial_2) \to (\mathcal{D}_1, \mathcal{D}_2\Lambda)$ as a correction that accounts for differences in $x^1$ and $x^2$ decay rates preventing (\ref{network}) from becoming a conservative dynamical system. In the same way that the ratio of decay rates (as measured by $-q$) can be viewed as a measure of how far the system has been distorted, the $q$-differential operators provide a way to re-scale activator/repressor concentrations and return to Hamiltonian form. In short, by working with $(\mathcal{D}_1, \mathcal{D}_2\Lambda)$ in place of $(\partial_1, \partial_2)$ one is effectively ignoring the difference in activator/repressor decay rates and performing calculations in the Hamiltonian limit ($\beta \to - \alpha$). In this way it is possible to interchange the two scenarios, Hamiltonian and deformed, calculating quantities in one picture before conveniently passing to the other.

Note that replacing differential operators with their $q$-analogues suggests it would be natural to move fully to the quantum plane $\mathbb{R}^2_q$. However, there is no obvious way to treat the non-commutativity of variables $(\hat{x}^1,\hat{x}^2)$ since this problem is never encountered when working with chemical concentrations. Nonetheless, if the same ordering as in (\ref{qHamil}) is maintained when passing to the quantum plane then the system (\ref{network}) has a natural Hamiltonian representation on $\mathbb{R}^2_q$

\begin{equation}
\dot{\hat{x}}^a = \sum_b \Omega^{ab} \hat{\partial}_b \hat{H} \ ,
\end{equation}
where

\begin{equation}
\hat{H} =  F_2(\hat{x}^1)- F_1(\hat{x}^2) - \beta \ \hat{x}^1 \hat{x}^2 \ ,
\end{equation}
 and the $F_a$ are understood to be expanded in terms of their arguments using (\ref{expand}). It follows that (\ref{network}) is conservative on $\mathbb{R}^2_q$ with Hamiltonian $\hat{H}$, Poisson bracket
 
\begin{equation}
\widehat{\{f,g\}} = \sum_{a,b} \Omega^{ab} \hat{\partial}_a f \hat{\partial}_b g \ ,
\end{equation}
and the total time derivative of any function $f:(\hat{x}^1, \hat{x}^2) \to \mathbb{R}$ given by

\begin{equation}
\dot{f} = \widehat{\{H,f\}} \ .
\end{equation}
Interestingly, this system is not invariant under the usual canonical nor $q$- canonical transformations due to non-commutativity of $(\hat{\partial}_1, \hat{\partial}_2)$. However, passage to the quantum plane may be viewed as a method of `$q$-Hamiltonisation' that enables one to work with (\ref{network}) as if it is a Hamiltonian system.  

As an example, the $q$-Hamiltonisation procedure can be applied to study a system previously used to examine multistability in the lactose utilization network of {\em Escherichia coli} \cite{Ozb}: 

\begin{eqnarray} \label{example} 
\tau_Y \dot{Y} =\frac{A}{1+R/R_0}-Y  \ , \\ \nonumber
\tau_X \dot{X} = BY - X \ ,
\end{eqnarray}
where

\begin{equation}
R = \frac{R_T}{1+(X/X_0)^n} \ .
\end{equation}
Here $X$ is the intracellular allolactose concentration, $Y$ is the concentration of the allolactose permease LacY, and $R$ the concentration of active repressor protein LacI (with total concentration $R_T$ and half-saturation concentration $R_0$). Allolactose binds to inhibit LacI, and so $R$ depends on the half-saturation concentration of allolactose binding ($X_0$) and a Hill coefficient $n$ that has been experimentally determined to be $n \approx 2$ \cite{Yag}.  It is clear that (\ref{example}) is of the form (\ref{network}) and therefore a $q$-deformed Hamiltonian system with $q=-\tau_Y/\tau_X$. Thus, $q$ is the negative ratio of allolactose and LacI depletion rates $\tau_X$ and $\tau_Y$, respectively, and the Hamiltonian limit $\tau_Y \to -\tau_X$ is equivalent to working with  Jackson derivatives and an $X$-dependent scaling operator in place of $(\partial_X, \partial_Y)$.

The authors of \cite{Ozb} studied bistability of solutions to (\ref{example}) under variation of $AB/\rho$ and $\rho=1+R_T/R_0$. Here $A$ is the maximal rate of LacY production, $B$ the allolactose uptake rate per LacY molecule, and $\rho$ is the repression factor that describes how tightly LacI is able to regulate gene expression. They were able to show that the region of bistability grows as the repression factor increases and is then determined by $AB/\rho$. It would be useful to try and explain bistability using $q$-Hamiltonisation with the understanding of how and what it means to work in the Hamiltonian limit. This is achieved by taking advantage of the fact that certain Hamiltonian systems can be solved implicitly. First, solving the  equation for $Y$ in terms of $X$ and $\dot{X}$ yields an inhomogeneous second-order equation of the form

\begin{equation}   
\ddot{X}+\frac{\tau_X+\tau_Y}{\tau_X \tau_Y} \dot{X} + \frac{1}{\tau_X \tau_Y}X = \frac{1}{\tau_X \tau_Y} \frac{AB}{1+R/R_0} \ ,
\end{equation}
which in the Hamiltonian limit reduces to

\begin{equation}
\ddot{X} = \frac{1}{\tau_X^2} \left[ X - \frac{AB}{1+R/R_0} \right] \equiv f(X) \ . \label{f}
\end{equation}
These equations have the general solution

\begin{equation} 
t+C_2=\pm \int \frac{dX}{\sqrt{2\int f(X) dX +C_1}} \ , \ \ \ \ C_1,C_2 \in \mathbb{R} \ ,
\end{equation}
and so the problem reduces to evaluating the integrals, but this is usually only possible for simple functions $f(X)$. For $f(X)$ in (\ref{f}) with arbitrary $n$ the integrand will involve the reciprocal square root of a hypergeometric function, but for the biologically relevant case $n=2$ one finds that 

\begin{equation}
2 \int f(X) dX + C_1 = \frac{1}{\tau_X^2}\left[ (X-AB)^2 + \frac{2AB}{\sqrt{\rho}} \frac{X_0 R_T}{R_0} \tan^{-1} \left(\frac{X}{X_0 \sqrt{\rho}}\right) +C_1' \right] \ ,
\end{equation}
where the constant of integration has been redefined. The region of bistability increases with $\rho$, and for large $\rho$

\begin{equation}
\frac{ab}{\sqrt{\rho}} \frac{X_0 R_T}{R_0} \tan^{-1} \left(\frac{X}{X_0 \sqrt{\rho}}\right) \sim  \frac{R_T}{R_0} \frac{AB}{\rho} X \ .
\end{equation}
Heightened bistability therefore implies (using the definition of $\rho$)

\begin{equation}
t+C_2 \sim \pm \tau_X \int \frac{dX}{\sqrt{(X-\frac{AB}{\rho} )^2+C_1''}} \ ,
\end{equation}
where once again the constant of integration has been redefined. Evaluating the integral, solving for $X$ explicitly in terms of $t$, and using initial conditions $X(0),\dot{X}(0)$ to fix the values of integration constants, one finds the Hamiltonian limit of the solution of greatest bistability to be

\begin{equation}
X \sim \frac{AB}{\rho} +\left(X(0)-\frac{AB}{\rho}\right) \cosh \left(\frac{t}{\tau_X}\right)+ \tau_X \dot{X}(0) \sinh \left(\frac{t}{\tau_X}\right) \ .
\end{equation}
How $AB/\rho$ controls bistability becomes immediately clear: its value dictates which initial conditions result in a solution that grows exponentially or is otherwise damped by determining the sign in front of the hyperbolic cosine. 

\section{Spatial coupling and the continuum limit}
The system (\ref{network}) describes a two-component feedback loop without reference to spatial dimensions. A natural extension of (\ref{network}) is to consider a series of $2N$ coupled equations describing a feedback loop whose constitutive molecules can travel throughout a tissue network or amongst different compartments of a cell \cite{deJ}. When modelling different compartments of a cell \cite{Gla} or multiple cells \cite{Bar}, diffusion of a molecular species between pairs of adjacent cells/compartments $i,j$ is assumed to occur proportionally to the difference in the concentration of that molecule between regions $i$ and $j$. When the spatial arrangement of $N$ compartment/cells is described by an $N \times N$ adjacency matrix $A$ whose only non-zero elements are $A_{ij}=1$ if region $i$ is connected to $j$, then a natural generalisation of (\ref{network}) is the coupled system of $2N$ equations

\begin{eqnarray} \label{coupled}
\dot{x}^i =f(y^i) - D_x \sum_j A_{ij}(x^i - x^j)  \ , \\ \nonumber
\dot{y}^i = g(x^i) - D_y \sum_j A_{ij}(y^i - y^j) \ ,
\end{eqnarray}
where $x^i,y^i$ are the concentrations of activator/repressor in region $i$ and $D_x,D_y$ are the diffusion coefficients for that molecule. The model (\ref{coupled}) can also be written as a $q$-deformed Hamiltonian system

\begin{equation}
\dot{z}^I = \sum_J \Omega_q^{IJ}\partial_J H \ \ \  I,J=1,...,2N \ , \label{largesystem}
\end{equation}
 where $z=(x^1,x^2,...,x^N,y^1,y^2,...,y^N)$ and

\begin{equation} 
\Omega_q =   \left( \begin{array}{cc} 0 &  \mathbb{I}_N \\   -q \mathbb{I}_N & 0 \end{array} \right) \ , \ \ \ q= - D_y/D_x \ ,
\end{equation}
with $\mathbb{I}_N$ the $N \times N$ identity matrix and

\begin{equation}
H= \sum_i  \left[ \int g(x^i) dx^i + \frac{D_y}{D_x} \int f(y^i) dy^i -D_y \sum_j A_{ij} (x^i-x^j)(y^i-y^j) \right] \ . \label{largehamil}
\end{equation}
From (\ref{largesystem}), a point $z_0 \in \mathbb{R}^{2N}$ is an equilibrium point of the coupled system (\ref{coupled}) if $\nabla H(z_0)=0$ . In the usual Hamiltonian picture this property combined with $\dot{H}=0$ means that $H$ or $-H$ is a good candidate for a Liapunov function provided one of them is positive definite for Liapunov's Theorem to apply. Then by Dirichlet's Theorem $z_0$ is a stable equilibrium if it is an isolated local maximum or local minimum, respectively, of the Hamiltonian $H$. However, in addition to this being a very strong condition, for the $q$-deformed case

\begin{equation}
\dot{H} = (1-q^{-1})\sum_i \partial_{x^i} H \partial_{y^i}H \neq 0 \ ,
\end{equation}
and so there is no guarantee that Dirichlet's Theorem holds. 

The above discussion means that one can not simply apply Dirichlet's Theorem to study the stability of the equilibrium points of (\ref{coupled}-\ref{largesystem}) even though these may be identified with zeros of $\nabla H$. However, proceeding via $q$-Hamiltonisation using the obvious choice of Jackson derivatives and scaling operators in place of  $(\partial_{x^i}, \partial_{y^i})$ is equivalent to working in the Hamiltonian limit $q \to 1$ where Dirichlet's Theorem applies. It is also instructive to see how $q$-Hamiltonisation operates by studying eigenvalues of the linearised version of (\ref{coupled}) in the same problem re-parameterised such that $z_0=0$. That is the linear system of equations

\begin{equation} \label{linearised}
 \left( \begin{array}{c}  \dot{\mbox{x}}  \\   \dot{\mbox{y}} \end{array} \right) = \left( \begin{array}{cc} -D_x \Delta &   f'(0)\times \mathbb{I}_N \\  g'(0)\times  \mathbb{I}_N & -D_y \Delta \end{array} \right) \left( \begin{array}{c}  \mbox{x}  \\   \mbox{y} \end{array} \right) \equiv M  \left( \begin{array}{c}  \mbox{x}  \\   \mbox{y} \end{array} \right) \ ,
\end{equation}
where $\mbox{x}=(x^1,...,x^N)$,  $\mbox{y}=(y^1,...,y^N)$ and $\Delta$ is the $N \times N$ Laplacian matrix associated to the graph with adjacency matrix $A$. By using the properties of block matrices in which all blocks commute \cite{Sil} eigenvalues $\lambda$ of $M$ must satisfy

\begin{equation}
\mbox{det}[M-\lambda \mathbb{I}_{2N}] = \mbox{det}[D_xD_y \Delta^2+ \lambda (D_x+D_y) \Delta + (\lambda^2- f'(0) g'(0)) \mathbb{I}_N] =0 \ .
\end{equation}
In the Hamiltonian limit $q \to 1$ ($D_y \to -D_x$) this simplifies to 

\begin{equation}
\mbox{det}[-D_x^2\Delta^2+(\lambda^2- f'(0) g'(0)) \mathbb{I}_N] =0 \ ,
\end{equation}
and therefore 

\begin{equation}
\Lambda ^2 = \frac{\lambda^2- f'(0) g'(0)}{D_x^2}
\end{equation}
is an eigenvalue of $\Delta^2$. From the general theory of eigenvalues of powers of operators and semi-positive definiteness of Laplacian matrices it follows that the $2N$ eigenvalues of $M$ in the Hamiltonian limit are

\begin{equation}
\pm \sqrt{D_x^2\Lambda_n^2+f'(0) g'(0)} \ , \ \ \ n= 1,2,...,N \ ,
\end{equation} 
where the $\Lambda_n$ are eigenvalues of $\Delta$. This reveals (in the Hamiltonian limit at least) how the stability of an equilibrium point of (\ref{coupled}) depends on the topology of the cellular/compartmental arrangement given by $A$ and the repressor/activator functions $f(y^i)$ and $g(x^i)$: There are many known upper bounds for the largest eigenvalue $\Lambda_N$ for the Laplacian of a connected graph. For example, Anderson and Morely \cite{And} showed that 

\begin{equation}
\Lambda_N \leq \mbox{max} \{ d_i+d_j :  \ A_{ij} \neq 0 \} \ ,
\end{equation}
where $d_i$ and $d_j$ are the degrees of vertices $i$ and $j$, respectively (i.e. number of neighbouring cells/compartments of regions $i$ and $j$). Thus, if $f'(0)$ and $g'(0)$ are of opposite sign with

\begin{equation}
|f'(0) g'(0)| > \mbox{max} \{ D^2_x(d_i+d_j)^2 : \ A_{ij} \neq 0 \} \ ,
\end{equation}
then in the Hamiltonian limit every eigenvalue of the linearised system (\ref{linearised}) becomes purely imaginary and the solution does not have exponentially growing terms. This corresponds to the statement that a two-component system containing one activator (e.g. mRNA) and one repressor (e.g. self-repressing protein) is stable provided that the number of adjacent cellular connections remains relatively small. 

As a final remark it is interesting to note that when $N \to \infty$ the system (\ref{coupled}) approaches a two-component reaction-diffusion system similar to that first considered by Turing \cite{Tur}. In this case $\mbox{x} \to \phi^1(t,\sigma)$ and $\mbox{y} \to \phi^2(t,\sigma)$, which are fields over some surface $\mathcal{S}$ parameterised by time and a continuous spatial parameter $\sigma$. The equations of motion for the fields then become

\begin{eqnarray} \label{fields}
\partial_t \phi^1 = f(\phi^2) - D_x \partial^2_{\sigma} \phi^1 \ , \\ \nonumber
\partial_t \phi^2 = g(\phi^1) - D_y \partial^2_{\sigma} \phi^2 \ .
\end{eqnarray}
In the $q \to 1$ limit these can be obtained from from a functional version of (\ref{largehamil}) using the usual formalism of classical Hamiltonian field theory.  Since there is no clear prescription for the functional analogue of a Jackson derivative however, the procedure for $q$-Hamiltonisation of such fields remains unclear. It will be left as an open problem for the time being.

\section{Concluding remarks}

As highlighted during the introduction, the problems encountered when attempting a mathematical description of biology appear to be partly associated with a lack of symmetry on the molecular scale. The importance of symmetry in physics is related to the existence of conserved quantities and the integrability of mechanical systems, something that dynamical biological processes do not generally share with their Hamiltonian counterparts. However, in this paper a model describing a large class of cellular processes was shown to admit a deformed version of Hamiltonian formalism, and reversing the deformation was demonstrated to reveal the analogue of a conserved quantity. In the particular cases considered, the nature of the deformation ($q$-deformation) is already well-understood, but it is likely that many other dynamical processes appearing in biology can be shown to reveal features of a novel deformed mechanics. One goal of mathematical biologists should be to identify and study these deformations so that non-conservative biological processes can be mapped to the classical formalism. There, many tools are available to advance our understanding of biological dynamics.

One distinctive feature of the models considered in this paper is that scale-invariance of molecular concentrations remains inherent even though general symplectic symmetry is deformed. Since scale-invariance is now a recognised property of many biological networks it is appropriate to ask the question whether allowed deformations will have scale-invariance at their heart, or if it is only the scale-invariant processes that can be returned to recognisable Hamiltonian form. The `$q$' for `quantum' is representative of the fact that $q$-deformations arose from deformation problems in quantum physics, but perhaps there is a larger set of `$b$'-deformations yet to be discovered and associated with scale-invariant biological processes. This type of speculation raises exciting possibilities for those working at the interface between physics and biology.
\\
\\

\section*{Acknowledgements}
I am grateful to J.M.V. Gomes for initiating discussions on $q$-deformations. I would like to acknowledge support from the Medical Research Council MC U105184332 and Peterhouse, Cambridge.





\end{document}